\documentclass[conference]{IEEEtran}
\IEEEoverridecommandlockouts

\usepackage{cite}
\usepackage{amsmath,amssymb,amsfonts}
\usepackage{algorithmic}
\usepackage{graphicx}
\usepackage{textcomp}
\usepackage{xcolor}
\usepackage{eso-pic}
\usepackage{enumitem}
\usepackage{tikz}
	
\usepackage{url}
\usepackage{lipsum}
\usepackage{pifont}

\usepackage{booktabs}

\usepackage{subcaption}

\newcommand{\PlaceIEEEPermNotice}{%
  \AddToShipoutPictureFG*{%
    \AtTextLowerLeft{%
      \raisebox{-0.55in}[0pt][0pt]{%
        \makebox[\textwidth][c]{%
          \begin{minipage}{0.97\textwidth}
          \centering
          \footnotesize
          \textcopyright~2026 IEEE. Personal use of this material is permitted.
          Permission from IEEE must be obtained for all other uses, in any current
          or future media, including reprinting/republishing this material for
          advertising or promotional purposes, creating new collective works, for
          resale or redistribution to servers or lists, or reuse of any copyrighted
          component of this work in other works.
          \end{minipage}%
        }%
      }%
    }%
  }%
}

\def\BibTeX{{\rm B\kern-.05em{\sc i\kern-.025em b}\kern-.08em
    T\kern-.1667em\lower.7ex\hbox{E}\kern-.125emX}}
\begin{document}

\title{Starlink Beacons for Passive LEO-Aided 9D Navigation}

\author{Nisanur~Camuzcu,~\IEEEmembership{Student Member,~IEEE,}
        Tiep~M.~Hoang,~\IEEEmembership{Member,~IEEE,}
        Alireza~Vahid,~\IEEEmembership{Senior~Member,~IEEE}% <-this % stops a space
        \thanks{Nisanur Camuzcu, Tiep M. Hoang and Alireza Vahid are with the Department of Electrical and Microelectronic Engineering, Rochester Institute of Technology, Rochester, NY 14623 USA (e-mails: nc1138@rit.edu; tmheme@rit.edu; arveme@rit.edu).
        
        This work was supported in part by NSF awards 2343964 and 2348589.}
        }

\maketitle
\PlaceIEEEPermNotice

\begin{abstract}
Global Navigation Satellite Systems (GNSS) underpin positioning, navigation, and timing (PNT), yet their low-power signals are easily blocked or disrupted, leaving gaps in PNT availability in contested environments (e.g., maritime) where interference, spoofing, or denial can occur. A key practical need is an independent, ubiquitous aiding signal that can be tracked passively and fused with inertial sensing to sustain full navigation-state estimation without dedicated or cooperative infrastructure. 
This paper presents an end-to-end LEO-aided hybrid framework that fuses GPS, Starlink downlink beacons, and an inertial measurement unit (IMU) in a 9D (3D position, 3D velocity, and 3D attitude) PNT system using an extended Kalman filter (EKF). We (i) extract Doppler-rate from Starlink downlink beacon tones by associating measurements with satellite IDs, (ii) benchmark beacon Doppler-rate against OFDM-derived range observables under a common processing/estimation pipeline, and (iii) integrate the resulting observable into inertial navigation. We evaluate GPS/IMU, Starlink/IMU, and GPS–Starlink–IMU using Fisher-information predictions, Monte Carlo simulations, and hardware measurements. Results show that Starlink Doppler-rate provides meaningful complementary PNT information, and can aid 9D estimation when GNSS is degraded or intermittently unavailable.
\end{abstract}


\section{Introduction}
\begin{figure} [ht]
    \centering
    \includegraphics[width=0.9\linewidth]{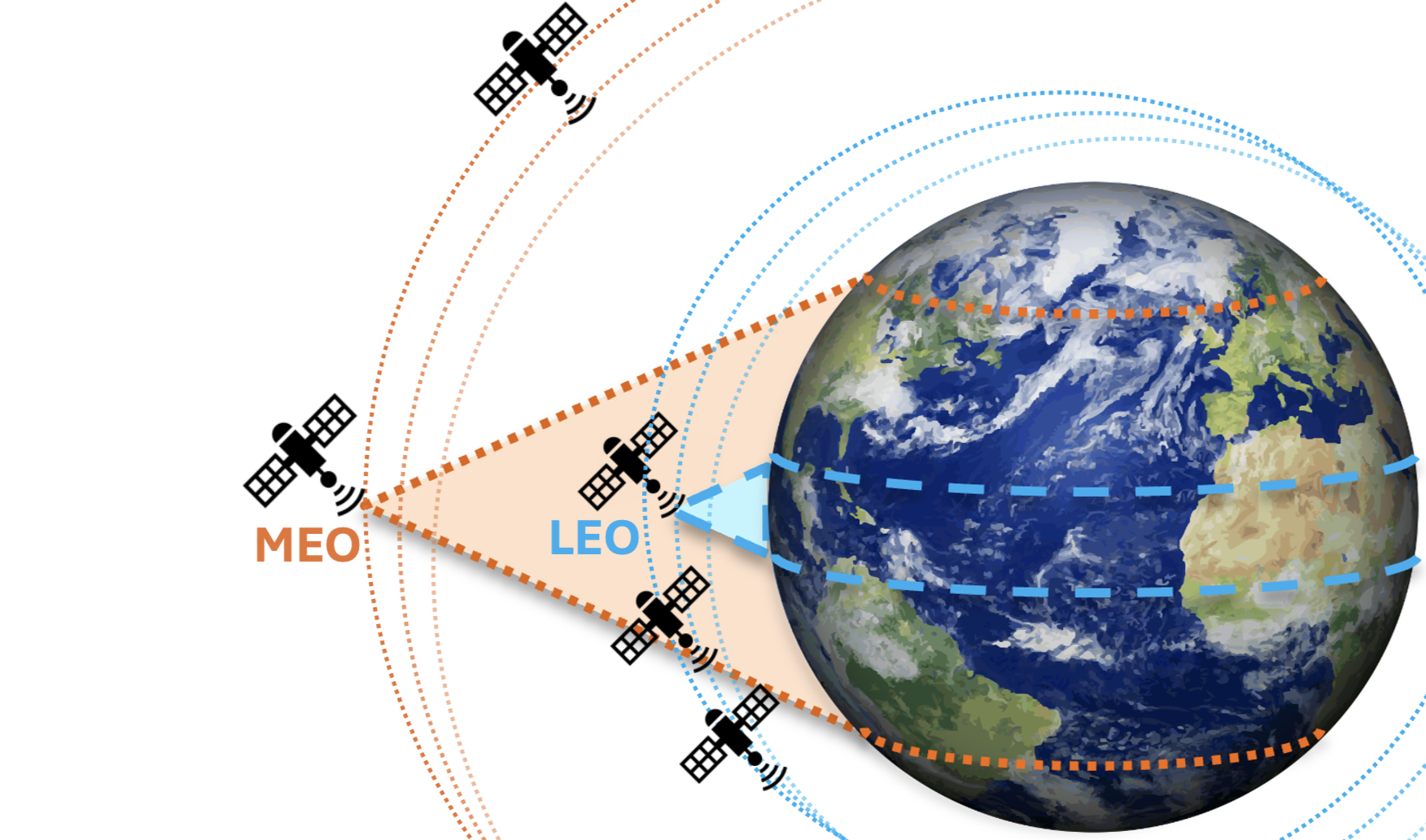}
    \vspace{-0.1 cm}
    \caption{LEO vs. MEO satellite orbits.}
    \label{fig:leo_meo}
    \vspace{-0.5 cm}
\end{figure}

The number of Low Earth Orbit (LEO) satellites is rapidly increasing to support broadband connectivity, driven by large-scale commercial constellations such as Starlink, OneWeb, and Kuiper~\cite{ref_pnt_leo}. While Global Navigation Satellite Systems (GNSS) remain the primary source of positioning data, their satellites operate in Medium Earth Orbit (MEO), resulting in higher latency and weaker signal strength compared to LEO counterparts. Fig.~\ref{fig:leo_meo} illustrates the relative orbital altitudes and coverage areas of LEO and MEO satellites. In contested maritime environments, GNSS signals are prone to interference, spoofing, and intermittent blockage from terrain, coastline, and ship structures~\cite{ref_deepsatloc}, motivating the need for more resilient positioning, navigation, and timing (PNT) solutions. Hence, the ubiquity of LEO satellite downlinks has positioned them as a promising complementary alternative to traditional GNSS.

LEO satellites support diverse missions, including broadband communication, Earth observation, and IoT connectivity~\cite{ref_pnt_leo}. Beyond enhancing their communication capabilities, recent research has focused on leveraging LEO satellites for the PNT applications. One of the most widely used LEO constellations is Starlink, with 9,621 satellites in orbit, 9,610 of them operational as of January 30, 2026~\cite{ref_starlink}. The scale, global coverage, and advanced communication payloads of such constellations make them attractive for complementing GNSS, particularly in challenging environments where traditional signals are degraded. Prior work has investigated LEO satellites for standalone PNT and for aiding/fusion with other systems (e.g., GNSS and inertial navigation system (INS))~\cite{ref_leo_pnt, ref_leo_gnss, ref_leo_ins}; yet, very few focus on 9D~\cite{ref_wang_9d,ref_9d_joint} or narrowband beacon tones rather than the full downlink waveform~\cite{ref_jardak_adam}. Although OFDM-derived observables are the most commonly studied, they can be intermittent for passive receivers in remote or low-traffic deployments with few concurrent active downlinks.

Exploiting this potential, this work investigates fusing global positioning system (GPS) pseudorange with Doppler rate from opportunistically detected Starlink beacons to localize a user device equipped with inertial measurement unit (IMU) and aims to improve \textit{localization dimensionality, accuracy, and robustness} in operationally contested settings where GPS may be degraded or denied. 
The system offers free global coverage with the ubiquity of Starlink satellites, where each transmits continuous-wave beacon signals in the Ku-band (10.7–12.7 GHz) downlink spectrum, primarily for terminal alignment. These beacon signals are broadcast by satellites as pure sine waves. As each strong beacon produces a high point in the FFT, the signals can be detected and the Doppler rate can be derived upon spectrum observations. 

\begin{table*}[!ht]
  \centering
    \vspace{-0.1 cm}
    \caption{LEO-based localization schemes (LEO-aided, 9D, or beacon-based).}
    \vspace{-0.1 cm}
  \resizebox{0.9\textwidth}{!}{
  \begin{tabular}{l|c|c|c|c|c}
    \toprule
    \textbf{Reference} 
      & \textbf{Signal Source} 
      & \textbf{LEO Measurements} 
      & \textbf{Sensors} 
      & \textbf{Algorithm} 
      & \textbf{Dimension} \\
    \midrule
    Wang et al. (2022)~\cite{ref_wang_9d} 
      & BS, LEO, GNSS 
      & Pseudorange, Doppler 
      & INS 
      & Fed. KF & 9D \\
      Jardak et al. (2023)~\cite{ref_jardak_adam}
      & LEO 
      & Doppler 
      & — 
      & EKF 
      & 3D \\
    Kassas et al. (2024)~\cite{ref_kassas} 
      & LEO, GNSS$^\textbf{*}$ 
      & Pseudorange, Doppler, Carrier phase 
      & IMU 
      & EKF 
      & 9D$^\textbf{*}$ \\
    Emenonye et al. (2024)~\cite{ref_9d_joint} 
      & BS, LEO 
      & ToA, Doppler 
      & — 
      & FIM 
      & 9D \\
    Kumar et al. (2024)~\cite{ref_deepsatloc} 
      & LEO, GPS (DL-aided)
      & AoA, Doppler 
      & IMU 
      & EKF 
      & 3D \\
    \textbf{Our work} 
      & \textbf{LEO, GPS} 
      & \textbf{Doppler rate} 
      & \textbf{IMU} 
      & \textbf{EKF} 
      & \textbf{9D} \\
    \bottomrule
  \end{tabular}
  \vspace{-0.1 cm}
  }
  \label{tab:comparison}

\noindent{\footnotesize $^{\textbf{*}}$GNSS is used only in the first 60 s. Although extended Kalman filter (EKF) uses 9D information, only 3D positioning results are presented.}
\vspace{-0.2 cm}
\end{table*}

Table~\ref{tab:comparison} summarizes representative related works, where LEO-aided navigation papers use range-type observables (time-of-arrival (ToA), pseudorange)~\cite{ref_wang_9d, ref_9d_joint, ref_kassas} and/or angle/phase observables (angle-of-arrival (AoA), carrier phase differences)~\cite{ref_deepsatloc, ref_kassas} that typically require known signal structure, tight synchronization, multi-antenna arrays, etc.; or LEO-only studies using beacons~\cite{ref_jardak_adam}. Also, these works primarily report 3D positioning performance~\cite{ref_jardak_adam, ref_deepsatloc}. Our work instead uses a lightweight, passive Starlink observable, Doppler-rate $\alpha$ from Ku-band central tone beacons, coupled with TLE-based association to resolve satellite IDs, and fuses $\alpha$ with IMU and GPS for 9D localization with a posterior-bound reference.

Building on prior LEO-assisted positioning studies, we provide the first end-to-end, theory-referenced benchmark that jointly \textbf{(i)} extracts Doppler-rate from Starlink beacons 
by associating measurements with satellite IDs, \textbf{(ii)} benchmarks Doppler-rate from beacons against OFDM-derived range observables under a common processing/estimation framework, \textbf{(iii)} incorporates the resulting Doppler-rate in a 9D inertial navigation filter (position, velocity, attitude) to enable passive, opportunistic LEO-aided positioning, and \textbf{(iv)} evaluates GPS+IMU, LEO+IMU, and LEO+GPS+IMU using posterior Cramér--Rao bound (PCRB), 
Monte Carlo simulations and hardware measurements in this work. Across the fusion modes, hardware broadly matches analytical/simulation trends. Moreover, the approach extends to other constellations given an observable deterministic downlink feature~\cite{ref_leo_pnt}.

\section{System Model}
In this section, we present our hybrid LEO-GPS positioning system to obtain position, velocity and orientation estimation in 3D. The objective is to obtain a 9D estimate, compare it with the theoretical lower bound, and validate it on hardware. For the LEO component, we focus on the Starlink constellation, exploiting its beacon signals (i.e., \textit{narrowband downlink central tones}) for PNT. By considering GPS/IMU and Starlink/IMU cases individually, as well as our proposed GPS-Starlink-IMU scenario, 
we demonstrate the advantage of fusing them together in contested environments. 

Satellite orbits are propagated using the Simplified General Perturbations model 4 (SGP4) algorithm. The SGP4 propagation uses two-line element (TLE) data, for all active Starlink and Navstar satellites on experimentation date as input. A satellite is considered visible if its elevation angle, measured from the user’s horizon, is above a threshold. This ensures that only satellites sufficiently high in the sky are used, avoiding those near the horizon that suffer from atmospheric distortion, multipath, and weak signals. We apply a 10° elevation mask for GPS satellites, consistent with common GNSS practice~\cite{ref_gps_elev}. Operating in the Ku-band and being more susceptible to atmospheric effects, multipath etc., 
Starlink user terminals typically employ a higher mask (i.e., 28°)~\cite{ref_starlink_elev}.

\begin{figure} [htbp]
    \centering
    \includegraphics[width=0.95\linewidth]{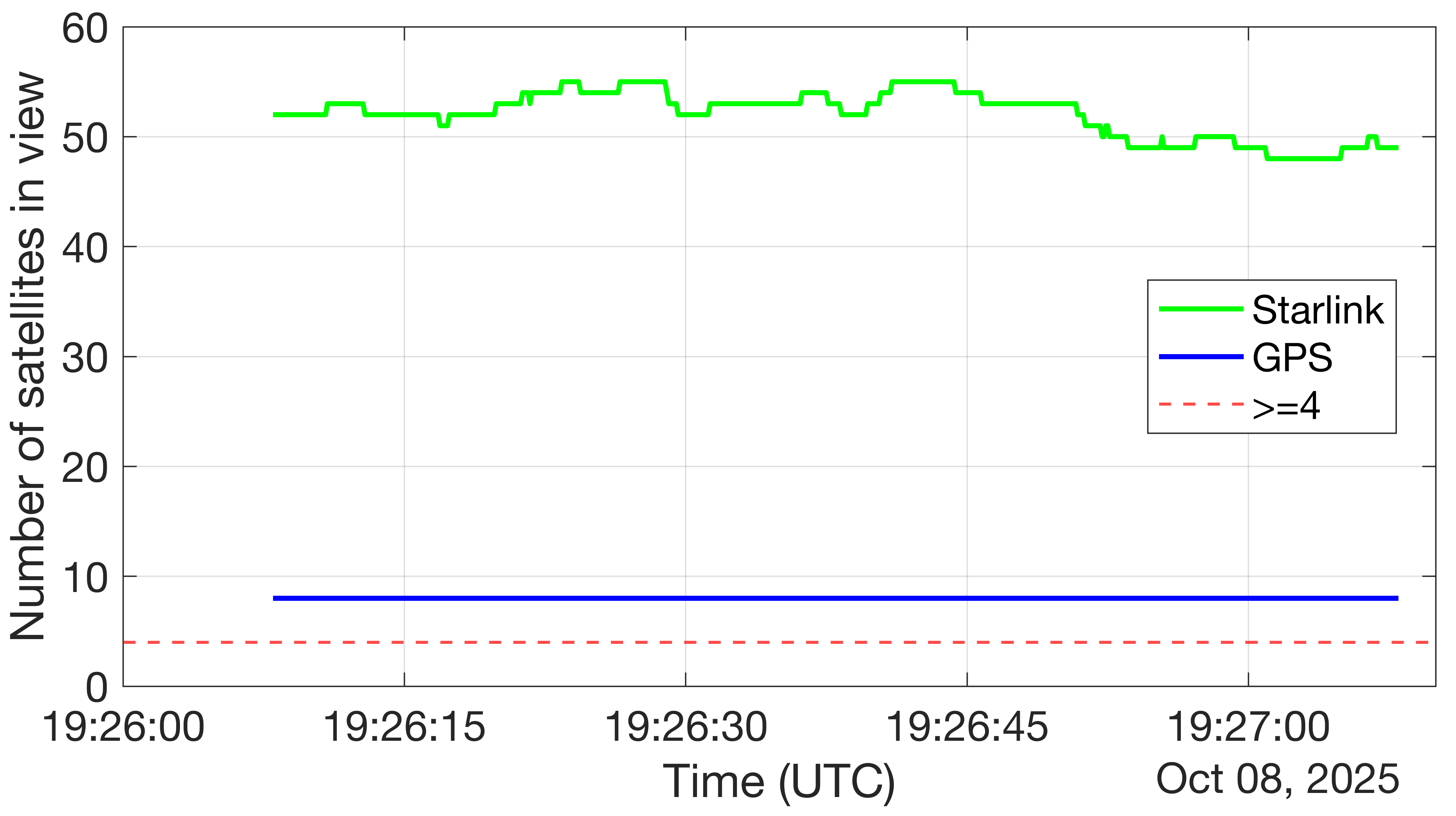}
    \vspace{-0.1 cm}
    \caption{Satellite observability from user position.}
    \vspace{-0.5 cm}
    \label{fig:visible_sat}
\end{figure}

We aim to localize a user, which is equipped with suitable receivers for satellite signals and a tactical-grade IMU. 
Navstar satellites provide pseudorange measurements, while for Starlink we model the pure sine-wave beacon, from which the satellite ID is matched and the Doppler rate is estimated. These measurements are then fused for 9D state estimation. For the selected date, satellite visibility is analyzed from the user’s perspective, as shown in Fig.~\ref{fig:visible_sat}. To ensure proper positioning, at least four Navstar satellites must provide pseudorange measurements, and at least four Starlink satellites must provide Doppler rate data as highlighted in Fig.~\ref{fig:visible_sat}.

\subsection{Starlink Beacons}

To use LEO satellites as reference points, the contributing spacecraft must first be identified. Unlike GNSS, Starlink beacons do not expose an explicit satellite identifier (e.g., PRN), and a raw spectrum or waterfall display does not directly reveal which satellite produced a given tone. Instead, identification exploits the fact that each satellite’s orbital motion induces a distinct Doppler signature at the receiver. Accordingly, we combine (i) TLE-driven Doppler rate (i.e., $\alpha$) predictions from simulation with (ii) measurements from captured signals~\cite{ref_star_beacon}.

\textbf{TLE-based prediction (simulation):}
For each candidate Starlink satellite $s$, SGP4/TLE propagation provides its Earth-Centered, Earth-Fixed (ECEF) position, velocity, and acceleration
$\bigl(\mathbf r_S(t),\mathbf v_S(t),\mathbf a_S(t)\bigr)$.
The receiver state is represented in a local East-North-Up (ENU) frame as
$\mathbf x(t)=[\mathbf p_{\text{enu}}^\top,\mathbf v_{\text{enu}}^\top,\boldsymbol\theta^\top]^\top$.
Using the known reference ECEF origin $\mathbf r_0$ and rotation $\mathbf R_{\text{enu}\to\text{ecef}}$,
\begin{equation}
\mathbf r_U(t)=\mathbf r_0 + \mathbf R_{\text{enu}\to\text{ecef}}\mathbf p_{\text{enu}}(t),
\qquad
\mathbf v_U(t)=\mathbf R_{\text{enu}\to\text{ecef}}\mathbf v_{\text{enu}}(t).
\end{equation}
The relative vectors are defined as
\begin{equation}
\mathbf r(t)=\mathbf r_S(t)-\mathbf r_U(t),\hspace{0.4 em}
\mathbf u(t)=\frac{\mathbf r(t)}{\|\mathbf r(t)\|},\hspace{0.4 em}
\mathbf v_{\text{rel}}(t)=\mathbf v_S(t)-\mathbf v_U(t).
\end{equation}
The Doppler rate (time derivative of Doppler shift) is modeled through the second derivative of range.
Following the kinematic identity used in our implementation, the range acceleration is approximated as
\begin{equation}
\ddot{\rho}(t)
\approx \underbrace{\mathbf u(t)^\top \mathbf a_S(t)}_{\text{radial sat. accel.}}
+ \underbrace{\frac{\|\mathbf v_{\text{rel}}(t)\|^2-\bigl(\mathbf u(t)^\top\mathbf v_{\text{rel}}(t)\bigr)^2}{\rho(t)}}_{\text{geometric term}},
\end{equation}
and the predicted Doppler rate is
\begin{equation}
\alpha_s(t)\triangleq \dot f_D(t)\approx -\frac{f_c}{c}\,\ddot{\rho}(t).
\end{equation}
This $\alpha_s(t)$ curve serves as the satellite-specific signature used for association and as LEO measurement model in the EKF.

\begin{figure} [htbp]
    \centering
    \includegraphics[width=\linewidth]{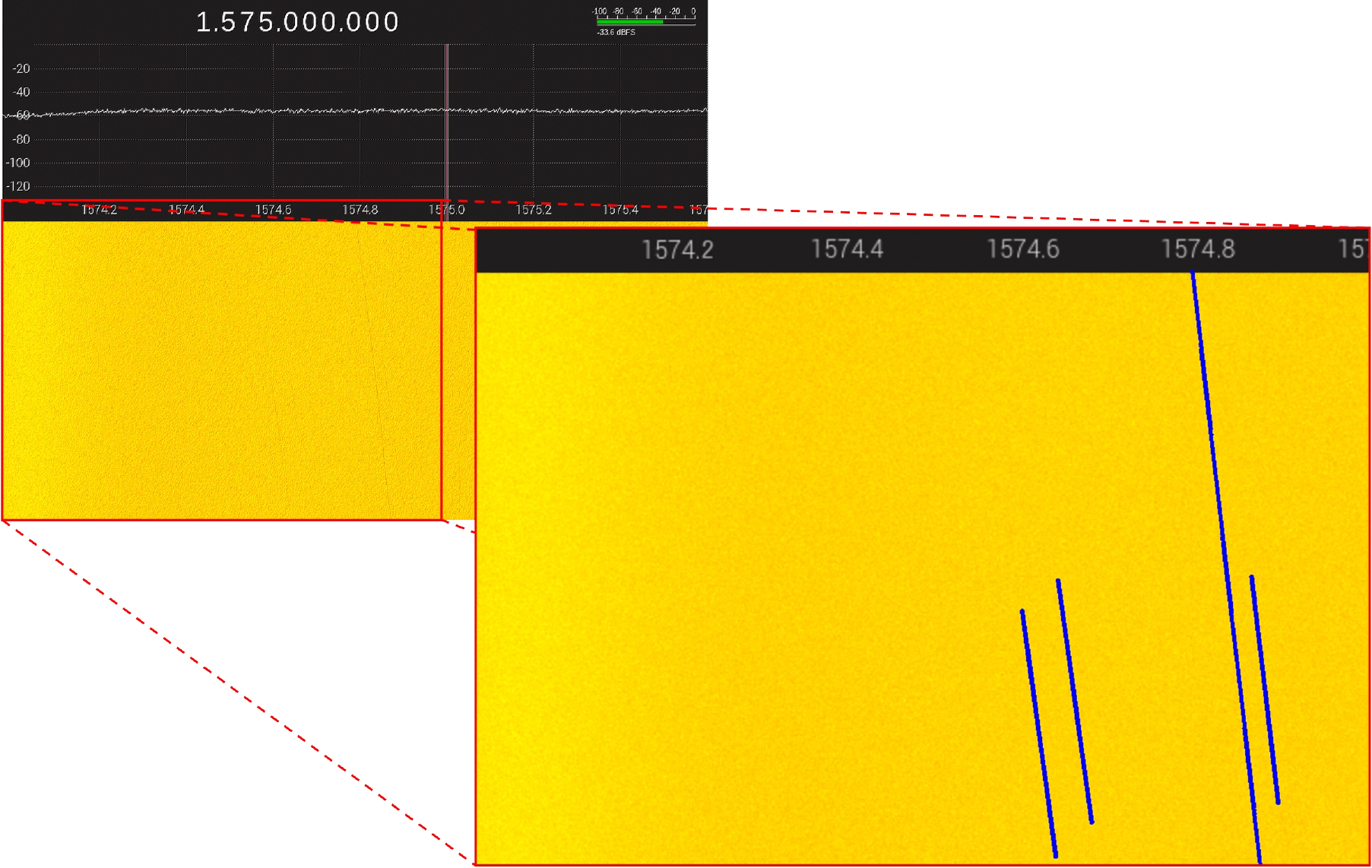}
    \caption{Beacon signals observed in the spectrum.}
    \vspace{-0.2 cm}
    \label{fig:spectrum}
\end{figure}

\textbf{Hardware measurement (from a spectrum capture):}
In practice, we often only observe a time-frequency trace (waterfall) where the beacon appears as a narrow ridge. From this observation, as in Fig.~\ref{fig:spectrum}, we extract a sequence of ridge points $\{(t_i,f_i)\}$ by tracking the peak-frequency bin of the tone in each time slice. This yields an estimate of the instantaneous Doppler trajectory $\hat{f}_D(t_i)=f_i-f_{\mathrm{ref}}$, where $f_{\mathrm{ref}}$ is the plot’s reference frequency. We then estimate the Doppler rate with 
\begin{equation}
\widehat{\dot{f}}_D \approx \frac{\hat{f}_D(t_{i+1})-\hat{f}_D(t_i)}{t_{i+1}-t_i},
\end{equation}

We match the measured Doppler signature to the TLE-predicted curves by selecting the satellite whose predicted $(f_D(t),\dot{f}_D(t))$ best matches the extracted ridge over the window and resolve the satellite ID for the observation~\cite{ref_star_beacon}. 

\textbf{OFDM ranging observable (simulation baseline):}
To provide a range-type LEO baseline in simulation, we define the OFDM baseline observable $d_i^{\rm ofdm}(t)$ as a differenced one-way range measurement with respect to a reference satellite:
\begin{equation}
d^{\text{ofdm}}_{i}(t)
\triangleq \bigl(\rho_i(t)-\rho_{\mathrm{ref}}(t)\bigr) + n^{\text{ofdm}}_{i}(t),
\label{eq:ofdm_diff_range}
\end{equation}
where $\rho_i(t)=\|\mathbf r_{S,i}(t)-\mathbf r_U(t)\|$ and 
$\rho_{\rm ref}(t)=\|\mathbf r_{S,\rm ref}(t)-\mathbf r_U(t)\|$ are the receiver-to-satellite ranges for satellite $i$ and the selected reference satellite at time $t$, respectively. 
Differencing cancels the receiver clock bias that would otherwise appear in a one-way time-of-arrival measurement, yielding a clock-free range-combination that depends primarily on geometry. In the EKF,~\eqref{eq:ofdm_diff_range} is used only as a simulated comparison baseline, with $n^{\text{ofdm}}_{i}(t)\sim\mathcal{N}(0,\sigma^2_{\text{ofdm}})$; the same measurement model is used when computing the reference lower-bound curves.

\subsection{Fusion and Lower Bound}

We estimate the navigation state by fusing (i) inertial measurements (accelerometer/gyroscope), (ii) GPS observables, and (iii) Starlink Doppler-rate measurements $\alpha$ extracted from Ku-band beacons. An extended Kalman filter (EKF) performs time propagation using the IMU, incorporates GPS and $\alpha$ through measurement updates at their respective time stamps.

To benchmark the achievable performance under the assumed process and measurement noise models, we compute a PCRB for the dynamic state~\cite{ref_pcrb}. We include the PCRB as a model-based lower bound to contextualize simulation scaling trends with the number of LEO satellites.

\section{Simulation and Experimental Setup}

\begin{figure} [b!]
    \centering     
    \vspace{-0.2 cm}  
    \includegraphics[width=\linewidth]{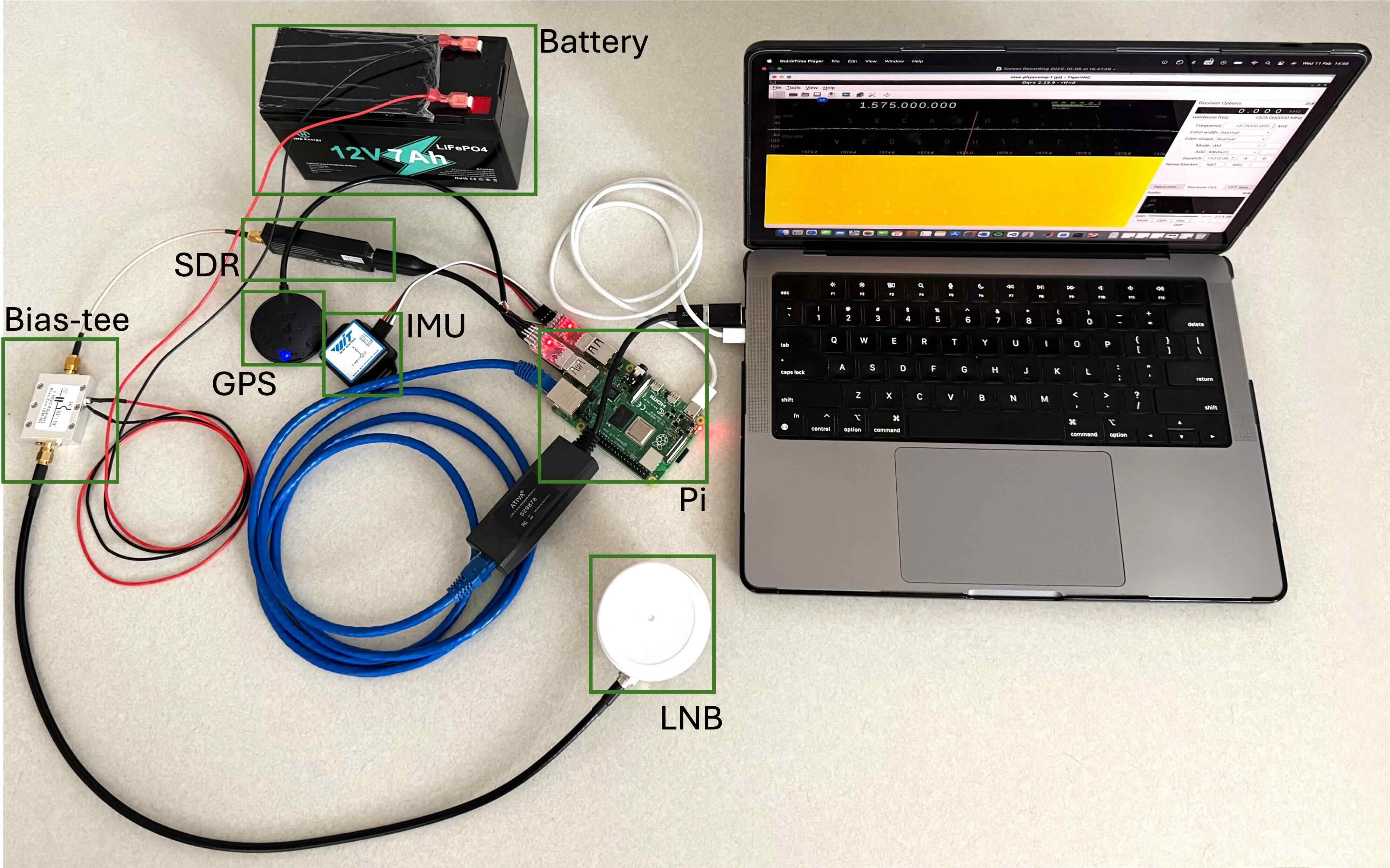}
    \caption{Hardware setup for measurements.}
    \vspace{-0.3 cm} 
    \label{fig:hw_setup}
\end{figure}

In this section, we present simulation setup followed by the used equipment in experimental setup.
\subsection{Simulation Setup}
To validate the proposed system, simulations were conducted in MATLAB R2024a. In the simulation, the user is assumed at a coordinate in Rochester, NY area (43.0848638, -77.6786127, 170), the location where hardware measurements are handled. During this period, the orientation is fixed to the north with $[\text{roll, pitch, yaw}] = [0,0,\pi/2]$). Obtaining the measurements over $20$~s, these information are fused in EKF to reach 9D estimation accordingly. Fig.~\ref{fig:visible_sat} demonstrates minimum number of visible Starlink satellites is 47, hence the simulation is conducted for the 5 to 45 satellites in that period.

\subsection{Hardware Setup}

Hardware measurements were collected outdoors on the RIT campus (Rochester, NY) for short-duration feasibility validation using the setup in Fig.~\ref{fig:hw_setup}.
During measurements, the GPS receiver, IMU, and LNB were placed in close proximity and treated as co-located; the separated layout in the figure is only for visual clarity. A Raspberry Pi 4 Model B collects time-synchronized sensor data from a u-blox NEO-M10-0-10 GNSS receiver and a WitMotion WT901C-TTL IMU, both configured at $10$~Hz and interfaced via UART/TTL. For satellite signal capture, a Ku-band low-noise block downconverter (LNB) feeds the RTL-SDR V3 front-end through a bias-tee. The LNB is powered from a $12$~V battery supply, which injects DC over the RF coax while the SDR receives the RF signal. Each spectrum capture covers a $20$~s period; hence all the Starlink, IMU, and GPS measurements within this interval are time-aligned and fused in the EKF. Due to capture/processing constraints, only five Starlink $\alpha$ updates are extracted and used for that duration.

\section{Results}
In this section, we present our simulation and experimental results with an overall performance analysis. We report both (i) a controlled simulation study that examines how localization performance scales with the number of LEO satellites and with different LEO observable types (Doppler-rate $\alpha$ versus a range-type baseline), and (ii) a short hardware capture using Ku-band Starlink beacons and a co-located GPS/IMU reference. Across both hardware and simulation, the results show that adding passive LEO observables can provide meaningful aiding, particularly in conditions where GPS is degraded or intermittently unavailable (e.g., contested maritime settings).

\begin{figure} [b!]
    \centering
    \vspace{-0.3 cm}
    \includegraphics[width=\linewidth]{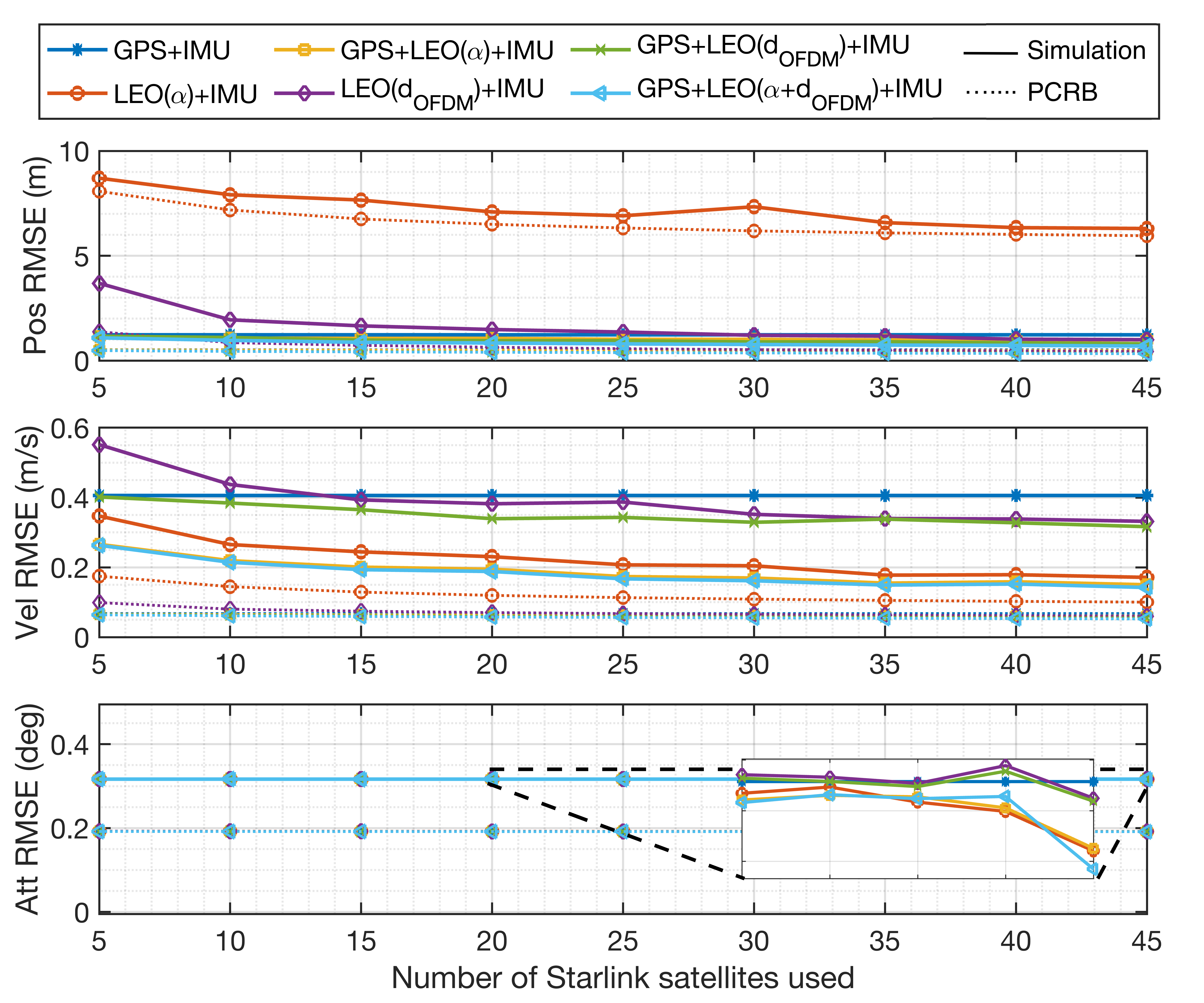}
    \caption{Performance in simulation and lower bound.}
    \label{fig:sim_rmse}
\end{figure}

\subsection{Simulation: Starlink Number and Observable Types}

Fig.~\ref{fig:sim_rmse} reports position/velocity/attitude RMSE versus the number of Starlink satellites used, along with a posterior lower bound curves (dashed) under the assumed process and measurement noise models. Across all cases, increasing the number of LEO satellites improves translational accuracy, reflecting improved geometry and measurement redundancy. For Doppler-rate, LEO($\alpha$)+IMU provides 6-9~m position RMSE; 
adding GPS reduces the RMSE toward the GPS-driven regime (1-2~m), but $\alpha$ alone does not match the accuracy of a range-type observable. For comparison, we include an \emph{OFDM range baseline} modeled as a \emph{differenced one-way range} measurement as in Eq. (\ref{eq:ofdm_diff_range}). As seen in Fig.~\ref{fig:sim_rmse}, range-type LEO observables improve position more strongly than $\alpha$, and combining $\alpha$ with range-type information yields the best translational performance among the LEO-aided cases. The attitude is relatively less sensitive to the LEO as it is primarily driven by inertial attitude dynamics provided by IMU.

\noindent \emph{\textbf{Takeaway}:} While range-based observables provide the strongest geometric constraint, \emph{passively} extracted Doppler-rate $\alpha$ remains a valuable \emph{fully passive} aiding measurement, especially in GNSS-degraded intervals, providing additional constraints that keep errors bounded and improve robustness as more Starlink satellites are incorporated.

\subsection{Hardware: Performance with passive Doppler-rate $\alpha$}

Satellite association quality was quantified by the post-fit Doppler-rate residual $|\Delta\alpha|$; across five matched beacons, the median $|\Delta\alpha|$ was 0.8~Hz/s (mean 1.46~Hz/s, max 3.4~Hz/s).

\begin{figure} [htbp]
    \centering
    \vspace{-0.3 cm}
    \includegraphics[width=\linewidth]{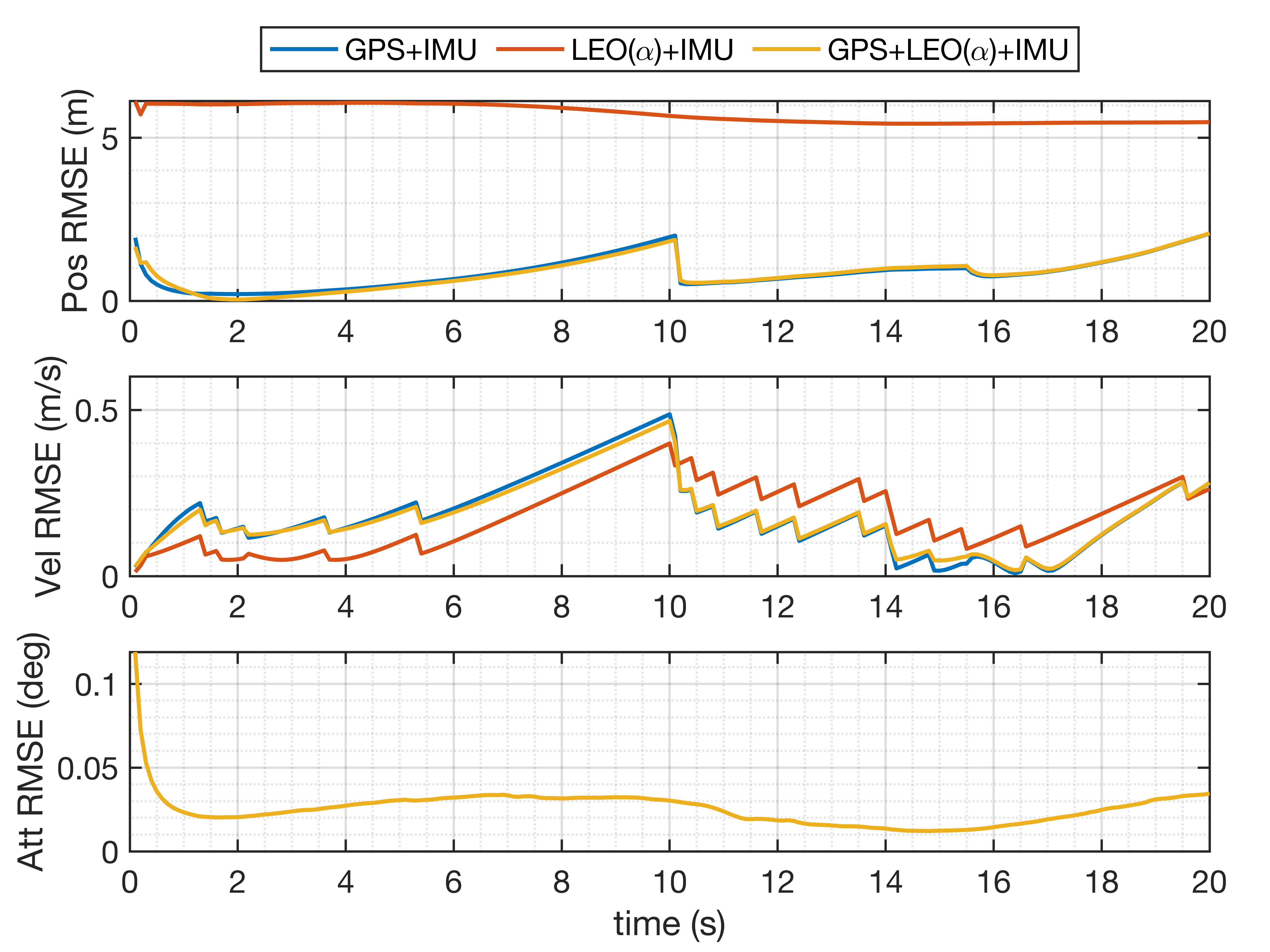}
    \caption{Performance with hardware measurements.}
    \label{fig:hw_rmse}
\end{figure}

Fig.~\ref{fig:hw_rmse} shows RMSE versus time for three modes in 9D EKF: GPS+IMU, LEO($\alpha$)+IMU, and GPS+LEO($\alpha$)+IMU. Using only passive Starlink Doppler-rate, LEO($\alpha$)+IMU yields a \emph{bounded} position error on the order of several meters and a stable velocity estimate, despite not using any GNSS pseudorange/carrier measurements. GPS+IMU achieves $\sim$1~m accuracy, and its error grows between strong updates as expected from inertial drift. The hybrid GPS+LEO($\alpha$)+IMU closely follows GPS+IMU when GPS is strong, while providing an additional independent observable that can help maintain bounded errors through GNSS transitions.

\noindent \emph{\textbf{Takeaway}:} Fully passive $\alpha$ measurements can support a practical fallback solution (bounded errors) and can be fused with GPS/IMU to improve robustness during GNSS degradation/transition periods, even when they do not dominate the solution under nominal GPS.

\begin{figure} [t!]
    \centering
    \includegraphics[width=\linewidth]{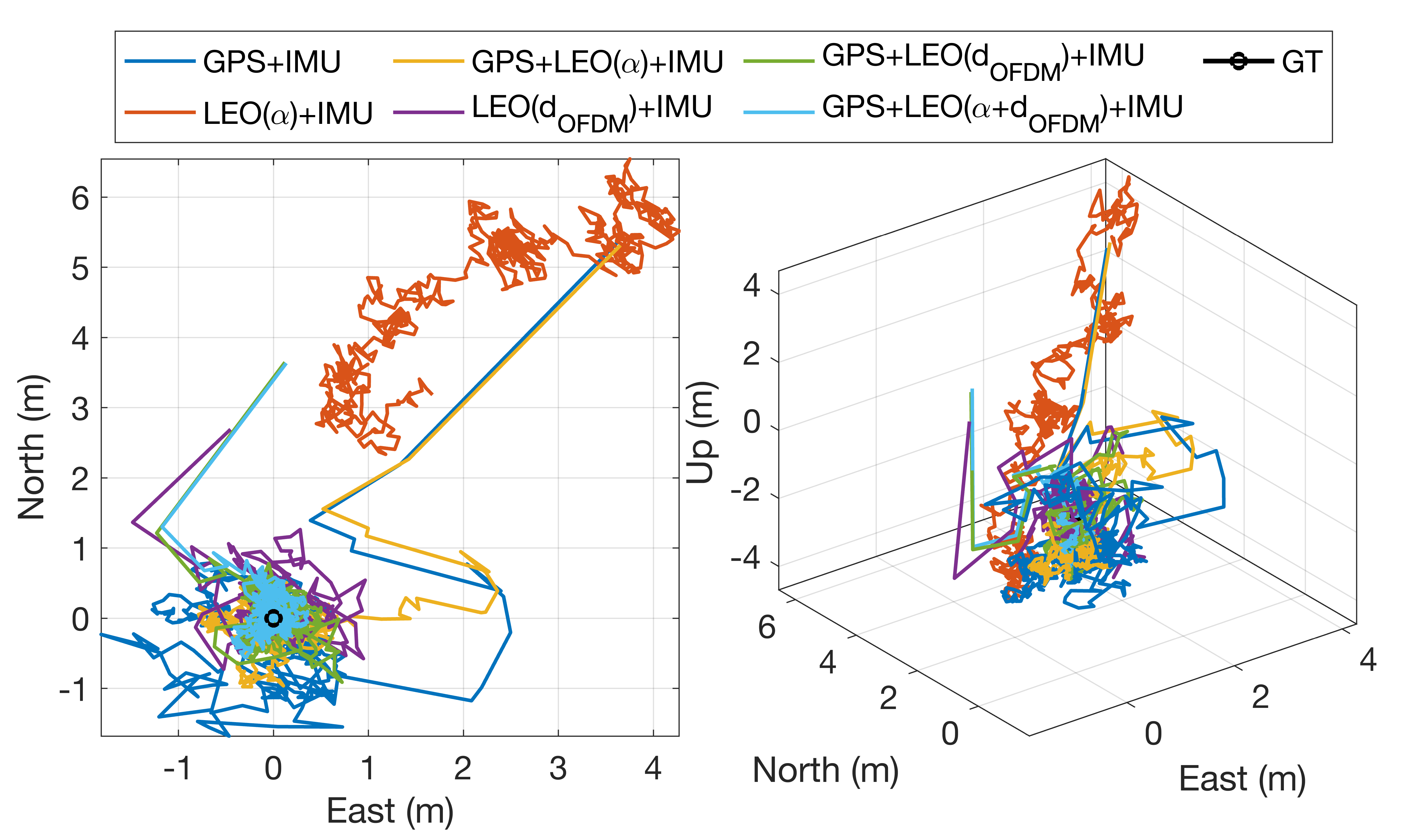}
    \vspace{-0.4 cm}
    \caption{2D-3D trajectories in simulation.}
    \vspace{-0.3 cm}
    \label{fig:sim_traj}
\end{figure}

Overall, in Fig.~\ref{fig:sim_rmse}, the Monte-Carlo RMSE follows the same qualitative scaling with the number of Starlink satellites and observables, and Fig.~\ref{fig:hw_rmse} demonstrates passive $\alpha$ measurements still yield bounded errors and stable velocity in practice. 

\noindent\emph{\textbf{Takeaway}:} Simulation and bound results explain the scaling trends, while the hardware results validate feasibility of passive $\alpha$ aiding under real beacon measurements.

\subsection{Trajectories in simulation and hardware}
\begin{figure} [t!]
    \centering
    \includegraphics[width=\linewidth]{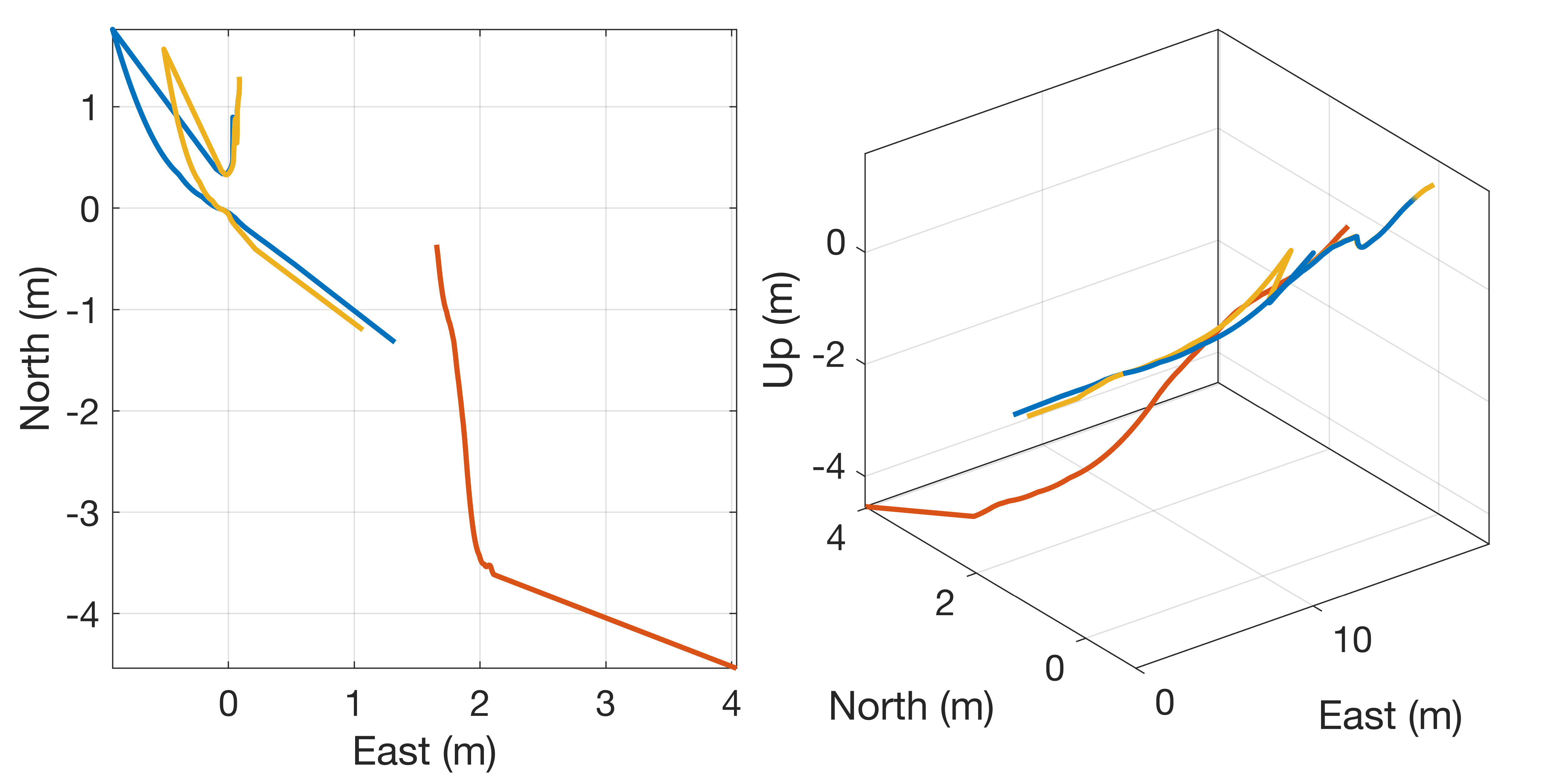}
    \vspace{-0.3 cm}
    \caption{2D-3D trajectories through hardware measurements.}
    \vspace{-0.3 cm}
    \label{fig:hw_traj}
\end{figure}

Fig.~\ref{fig:sim_traj} and Fig.~\ref{fig:hw_traj} visualize representative ENU trajectories, where GT denotes ground truth. In both settings, LEO($\alpha$)+IMU produces a trajectory that can drift relative to the ground truth, consistent with its RMSE behavior, whereas GPS-aided solutions remain concentrated near the reference. The hybrid GPS+LEO($\alpha$)+IMU trajectory closely overlaps the GPS+IMU trajectory, indicating that $\alpha$ updates act primarily as a robustness aid but can also help when GPS fails. Notably, the short hardware interval can appear slightly better than the simulation curves due to: (i) conservative assumptions in simulation and (ii) more favorable GPS/IMU performance and motion during the capture.

\noindent \emph{\textbf{Takeaway}:} The trajectory plots reinforce the RMSE trends: $\alpha$-only yields bounded but biased/drifting translation, while hybrid fusion preserves GPS-level accuracy when available and improves resilience when GPS quality degrades.

\section{Conclusion}
We demonstrated an end-to-end 9D navigation pipeline that passively extracts Starlink beacon Doppler-rate, associates measurements to satellites using TLE-predicted Doppler signatures, and fuses $\alpha$ Doppler rate with IMU and GPS in an EKF. Across simulation and hardware, Starlink Doppler-rate provides a stable sub-10~m positioning error and a reliable velocity estimation. In GPS-degraded conditions, \emph{passively} extracted Starlink Doppler-rate provides a practical fallback observable, and fusing GPS-LEO($\alpha$)-IMU improves 9D robustness by keeping errors bounded through GNSS dropouts and recoveries. More broadly, the concept generalizes to other LEO constellations whenever a trackable deterministic downlink component (e.g., beacon/sync/pilot) is observable with suitable receiver hardware~\cite{ref_leo_pnt}. 
Simulation trends further show that increasing the number of LEO satellites and incorporating range-type LEO observables can improve translational accuracy, motivating future work on more measurement-rich downlink processing and longer-duration field tests.

\end{document}